\documentclass[twocolumn,aps,10pt,prl]{revtex4-1}
\def\bwt{\begin{widetext}}

\def\ewt{\end{widetext}}

\def\be{\begin{equation}}

\def\ee{\end{equation}}

\def\bea{\begin{eqnarray}}

\def\eea{\end{eqnarray}}

\def\bean{\begin{eqnarray*}}

\def\eean{\end{eqnarray*}}

\def\bary{\begin{array}}

\def\eary{\end{array}}

\def\bit{\begin{itemize}}

\def\eit{\end{itemize}}

\newcommand{\dd}{\mathrm{d}}

\usepackage{color}
\usepackage{graphicx}
\usepackage{hyperref}
\usepackage{times}

\begin{document}

\title{The Minimal GUT with Inflaton and Dark Matter Unification}

\author{Heng-Yu Chen$^{1}$}
\author{Ilia Gogoladze$^{1}$}
\author{Shan Hu$^{2}$}
\author{Tianjun Li$^{3,4}$}
\author{Lina Wu$^{3,5}$}
\affiliation{
$^1$Bartol Research Institute, Department of Physics and Astronomy,
University of Delaware, Newark, DE 19716, USA
}
\affiliation{
$^2$Department of Physics, Hubei University, Wuhan 430062, P. R. China
}

\affiliation{
$^3$Key Laboratory of Theoretical Physics, Institute of Theoretical Physics,
Chinese Academy of Sciences, Beijing 100190, China
}
\affiliation{
$^4$School of Physical Sciences, University of Chinese Academy of Sciences,
No.~19A Yuquan Road, Beijing 100049, China
}
\affiliation{
$^5$School of Physical Electronics, University of Electronic Science and Technology of China,
Chengdu 610054, P. R. China
}

%\date{today}

%%%%%%%%%%%%%%%%%%%%%%%%%%%%%%%%%%%%%%%%%%%%%%%%%%%%%%%%%%%%%%%%%%%%%%%%%%%%

\begin{abstract}

Giving up the solutions to the fine-tuning problems, we propose
the non-supersymmetric flipped $SU(5)\times U(1)_X$ model based on the
minimal particle content principle,  which can be constructed from
the four-dimensional $SO(10)$ models,
five-dimensional orbifold $SO(10)$ models, and local F-theory $SO(10)$ models.
To achieve gauge coupling unification, we introduce one pair of vector-like fermions,
which form complete $SU(5)\times U(1)_X$ representation.
 Proton lifetime is around $5\times 10^{35}$ years,
 neutrino masses and mixing  can be explained via seesaw mechanism,
baryon asymmetry can be generated via leptogenesis, and vacuum stability problem can be solved
as well. In particular, we  propose that inflaton and dark matter particle
can be unified to a real scalar field with $Z_2$ symmetry, which is not an axion and does not
have the non-minimal coupling to gravity. Such kind of scenarios can be applied to the
generic scalar dark matter models. Also, we find that the vector-like particle corrections
to the $B_s^0$ masses can be about 6.6\%, while their corrections to the $K^0$ and $B_d^0$
masses are negligible.

\end{abstract}
\maketitle

%\section{Introduction}

{\bf Introduction~--}~It is well-known that a Standard Model (SM) like Higgs boson $(h)$ with
mass $m_h=125.09\pm 0.24$~GeV was discovered at the LHC~\cite{ATLAS, CMS, moriond2013},
and thus the SM particle content has been confirmed.
Moreover, there are many possible directions for new physics beyond the SM:
supersymmetry, extra dimensions, strong dynamics or say composite Higgs field, extra gauge symmetries,
and Grand Unified Theory (GUT), etc. However,
 we do not have any new physics signal at the 13 TeV Large Hadron Collider (LHC) yet.
Therefore, we may need to reconsider the principle for new physics beyond the SM,
and then propose the promising models.

First, let us briefly review the convincing evidence for new physics beyond the SM

\noindent $\bullet$ Dark Matter (DM) is a necessary ingredient of
cosmology, considering  the cosmic microwave background (CMB) or   the rotation curves
of spiral galaxies, etc~\cite{Spergel:2003cb, Ade:2015xua}.

\noindent $\bullet$ Dark Energy (DE) is required due to the concordance of data from
cosmic microwave anisotropy \cite{Spergel:2003cb}, galaxy clusters
(see, {\it e.g.}\/, \cite{Verde:2001sf}), and high-redshift Type-IA
supernovae \cite{Perlmutter:1998np,Riess:1998cb}.

\noindent $\bullet$ The non-zero  masses and  mixing of neutrinos from the
atmospheric \cite{Fukuda:1998mi} and solar
neutrino experiments \cite{Ahmed:2003kj}, as well as the reactor anti-neutrino
experiments \cite{Eguchi:2003gg}, etc.

\noindent $\bullet$ A larger fraction of baryonic matter compare to  anti-matter
in the Universe, {\it i.e.}, the cosmic baryon asymmetry $\eta = n_B/n_{\gamma} =
6.05 \pm 0.07 \times 10^{-10}$~\cite{Ade:2015xua}.

\noindent $\bullet$ The nearly scale-invariant, adiabatic, statistically isotropic,
 and Gaussian density fluctuations (see, {\it e.g.}\/,
\cite{Komatsu:2003fd}) point to cosmic inflation, which can solve the
horizon and flatness problems of the Universe as well.

Second, there are two kinds of theoretical problems in the SM: fine-tuning problems
and aesthetical problems. The fine-tuning problems are:
(i) The cosmological constant problem: why the cosmological constant is so tiny?
(ii) The gauge hierarchy problem: the SM Higgs boson mass square is not stable against
quantum corrections and has quadratic divergences, while the electroweak scale is
about 16 order smaller than the reduced Planck scale
$M_{\rm Pl}\simeq 2.43\times 10^{18}~{\rm GeV}$; (iii) The strong CP problem:
the $\theta$ parameter of the Quantum Chromodynamics (QCD) is smaller than
$10^{-10}$ from the measurements of
the neutron electric dipole moment~\cite{Afach:2015sja, Schmidt-Wellenburg:2016nfv};
(iv) The SM fermion mass hierarchy problem: electron mass is about 5 order smaller than
top quark mass. Also, the aesthetic problems are: (i) No explanation for the structure of gauge interactions; 
(ii) No explanation of fermion mass structures; (iii) No explanation for charge quantization; (iv) No realization of gauge coupling unification. The aesthetic problems can be solved
in the Grand Unified Theories (GUTs) if we can
realize gauge coupling unification. In addition, the SM Higgs quartic coupling
becomes negative around $10^{9}$ GeV for central measured values of the SM parameters.
Thus, the SM Higgs vacuum  is not stable, which is called
stability problem~\cite{Degrassi:2012ry, Tang:2013bz, Buttazzo:2013uya}.
Interestingly, the measured Higgs mass roughly corresponds to the minimal
values of the Higgs quartic and top Yukawa coupling as well as the maximal values of
the SM gauge couplings allowed by vacuum meta-stability~\cite{Buttazzo:2013uya}. In short, the SM
vacuum can be meta-stable while not absolutely stable.

Neglecting the fine-tuning and aesthetic problems,
Davoudiasl, Kitano, Li and Murayama proposed the New Minimal
Standard Model (NMSM) to address the above new physics evidence
based on the principle of the minimal particle content
and most general renormalizable Lagrangian~\cite{Davoudiasl:2004be}.
The dark energy is explained by a tiny cosmological constant,
dark matter particle is a real scalar with $Z_2$ symmetry, inflaton is another real scalar,
neutrino masses and mixing can be addressed via seesaw mechanism~\cite{Seesaw},
and baryon asymmetry is generated via leptogenesis~\cite{Fukugita:1986hr}.
Interestingly, inflation is still consistent with current observations
if we consider the polynomial inflation~\cite{Li:2014zfa},
and the NMSM is still fine via  meta-stability due to the minimality principle.
Later, Asaka, Blanchet and Shaposhnikov proposed the $\nu$MSM to explain baryon asymmetry,
neutrino oscillations, and dark matter via sterile right-handed neutrinos
with masses around a few KeV~\cite{Asaka:2005an, Asaka:2005pn}.
In 2015, Salvio proposed a simple SM completion~\cite{Salvio:2015cja}.
Compared to the NMSM, the main differences are: (i) The dark matter candidate is
axion, and the strong CP problem is solved via the invisible axion model proposed
by Kim, Shifman, Vainshtein, and Zakharov (KSVZ)~\cite{KSVZ}; (ii) Higgs field as the inflaton.
On the other hand,  the string landscape can explain the cosmological constant problem
and gauge hierarchy problem~\cite{String, Weinberg},
but cannot explain the strong CP problem~\cite{Donoghue:2003vs}. However,
for the non-supersymmetric KSVZ model, at least it is not clear whether the
string landscape can stabilize the axion. This is the reason why Barger, Chiang, Jiang and Li
proposed the intermediate-scale supersymmetric KSVZ axion model~\cite{Barger:2004sf}.
Also, there exists a serious difficulty for Higgs inflation since the scale of the 
Higgs field during inflation is larger than that of the perturbative unitarity 
violation~\cite{Burgess:2009ea, Barbon:2009ya}.
Recently, Ballesteros, Redondo, Ringwald and Tamarit proposed
the SM Axion Seesaw Higgs portal inflation (SMASH)
model~\cite{Ballesteros:2016euj, Ballesteros:2016xej}
to explain the above new physics evidence and strong CP problem, where the
axion is dark matter candidate, as in Ref.~\cite{Salvio:2015cja}.

In this paper, we still neglect the fine-tuning problems, and study
the Minimal GUT which can solve all the aesthetic problems in the SM.
We consider the non-supersymmetric flipped $SU(5)\times U(1)_X$ models~\cite{smbarr, dimitri, AEHN-0}.
To achieve gauge coupling unification, we introduce one pair of vector-like fermions,
which form complete $SU(5)\times U(1)_X$ representation.
This kind of models can be constructed in the four-dimensional $SO(10)$ models~\cite{Huang:2006nu},
five-dimensional orbifold $SO(10)$ models~\cite{Barr:2002fb}, and
local F-theory $SO(10)$ models~\cite{Jiang:2008yf, Jiang:2009za}.
The doublet-triplet splitting problem can be solved at tree level,
 proton lifetime is about $5\times 10^{35}$ years,
 neutrino masses and mixing can be explained via seesaw mechanism, baryon asymmetry can be generated
via leptogenesis, and stability problem can be solved as well.
Especially, we for the first time show that inflaton and dark matter particle can be unified
to a real scalar field with $Z_2$ symmetry, unlike all the previous models where
such scalar is either an axion or has non-minimal coupling to
gravity (Ricci scalar $R$)~\cite{Salvio:2015cja, Ballesteros:2016euj, Ballesteros:2016xej, Dias:2014osa,
Tenkanen:2016twd, Daido:2017wwb}. In other words, this is a brand new unification of
the inflaton and dark matter particle. After inflation, the interaction between
inflaton and Higgs field is reduced to that in the NMSM. Thus, such kind of scenarios
can be applied to the general scalar dark matter models. Furthermore,
we find that the corrections to the $B_s^0$ masses from vector-like particles
can be about 6.6\%, while their corrections to the $K^0$ and $B_d^0$ masses are negligible.

{\bf Model Building~--}~We introduce three families of the SM fermions, two Higgs fields $H$ and $h$,
and one pair of vector-like particles ($F_x$, $\overline{F}_x$), whose quantum numbers under
the $SU(5)\times U(1)_{X}$ gauge group and SM particle contents are
\bea
&&F_i/\Phi/F_x={\mathbf{(10, 1)}},~ {\bar f}_i={\mathbf{(\bar 5, -3)}},~
{\bar l}_i={\mathbf{(1, 5)}},~ \nonumber \\
&&\phi={\mathbf{(5, -2)}},~\overline{F_x}={\mathbf{(\overline{10}, -1)}}.~\nonumber \\
&&F_i=(Q_i, D^c_i, N^c_i),~{\overline f}_i=(U^c_i, L_i),~{\overline l}_i=E^c_i,~\nonumber\\
&&\Phi=(Q_\Phi, D_\Phi^c, N_\Phi^c),~\phi=(D_\phi, H),~\nonumber\\
&& F_x=(Q_x, D_x^c, N_x^c),~\overline{F}_x=(Q_x^c, D_x, N_x)~,~
\label{Particle-Content}
\eea
where $i=1, 2, 3$, and $Q_i$, $L_i$, $U_i^c$, $D_i^c$,  $L_i$,  $E_i^c$, $N_i^c$, and $H$
are the left-handed quark and lepton doublets, right-handed up-type quarks,
down-type quarks, charged leptons, neutrinos, and Higgs field, respectively.

To break the $SU(5)\times U(1)_{X}$ gauge symmetry down to the SM
gauge symmetry, we introduce the following Higgs potential at the GUT scale
\bea
{V}_{\rm GUT}&=& \lambda_\Phi (|\Phi|^2-v^2_{\rm GUT})^2 + \lambda|\epsilon^{ijklm} \Phi_{kl} \, \phi_m|^2~,~\,
\label{pgut}
\eea
where $i,~j,~k,~l,$ and $m$ are $SU(5)$ Lie Algebra indices.
After minimizing the potential, the $\Phi$  field acquires a Vacuum Expectation Value (VEV)  
at   $<N^c_\Phi>=v_{\rm GUT}$  component, and then the $SU(5)\times U(1)_X$  gauge symmetry is broken 
down to the SM gauge symmetry. As the result,  $Q_{\Phi}$ and imaginary component 
of $N^c_{\Phi}$ are eaten by superheavy gauge bosons,
while $D_\Phi^c$ and real component of $N^c_{\Phi}$ acquire GUT-scale masses.
The last term in Eq.~(\ref{pgut}) will generate  GUT-scale mass to  $D_\phi$ 
but not the SM Higgs doublet $H$. Thus, we naturally obtain the doublet-triplet splitting
at tree level, but we do need fine-tuning to keep the doublet light due to quantum corrections.

The Yukawa coupling and vector-like mass terms are
\bea
-{\cal L} &=& M_V F_x \overline{F}_x + \mu_i F_i \overline{F}_x +
y_{ij}^{D} F_i F_j \phi + y_{ij}^{U \nu} F_i  {\overline f}_j {\overline \phi}
\nonumber \\  &&
+ y_{ij}^{E} {\overline l}_i  {\overline f}_j \phi+ y_x^{D} F_x F_x \phi
+ y^{D}_{xi} F_x F_i \phi
+ y_{xi}^{U \nu} F_x  {\overline f}_i {\overline \phi}
\nonumber \\  &&
+y_{ij}^N \frac{1}{M_{\rm Pl}} \overline{\Phi} \, \overline{\Phi} F_i F_j
+ y_x^{N} \frac{1}{M_{\rm Pl}} \overline{\Phi}\,  \overline{\Phi} F_x F_x
\nonumber \\  &&
+y_{xi}^{N} \frac{1}{M_{\rm Pl}} \overline{\Phi}\, \overline{\Phi} F_x F_i
+ y_x^{\overline{N}} \frac{1}{M_{\rm Pl}} {\Phi} {\Phi} \overline{F}_x \overline{F}_x
+ {\rm H.C.}~,~
\label{Yukawa}
\eea
where $M_{\rm Pl}$ is the reduced Planck scale.
Once $\Phi$ field develops a VEV, the   $N^c_i$, $N^c_x$, and $N_x$  can obtain masses  
around $10^{14}$ GeV times their corresponding Yukawaw couplings.
 Assuming $M_V \approx  1$~TeV and $ \mu_i \approx 0$~TeV, 
we can have the vector-like particles  $(Q_x,~Q_x^c)$ and $(D_x,~ D_x^c)$ at low energy
without involving any more fine tuning.  As shown previously, this particle content 
leads to the gauge coupling
unification ~\cite{Amaldi:1991zx, Gogoladze:2010in}. The main difference is that
these vector-like particles in our models form the complete GUT multiplets, which is 
an interesting point as well.

\begin{figure}
  \centering
  \includegraphics[width=\columnwidth]{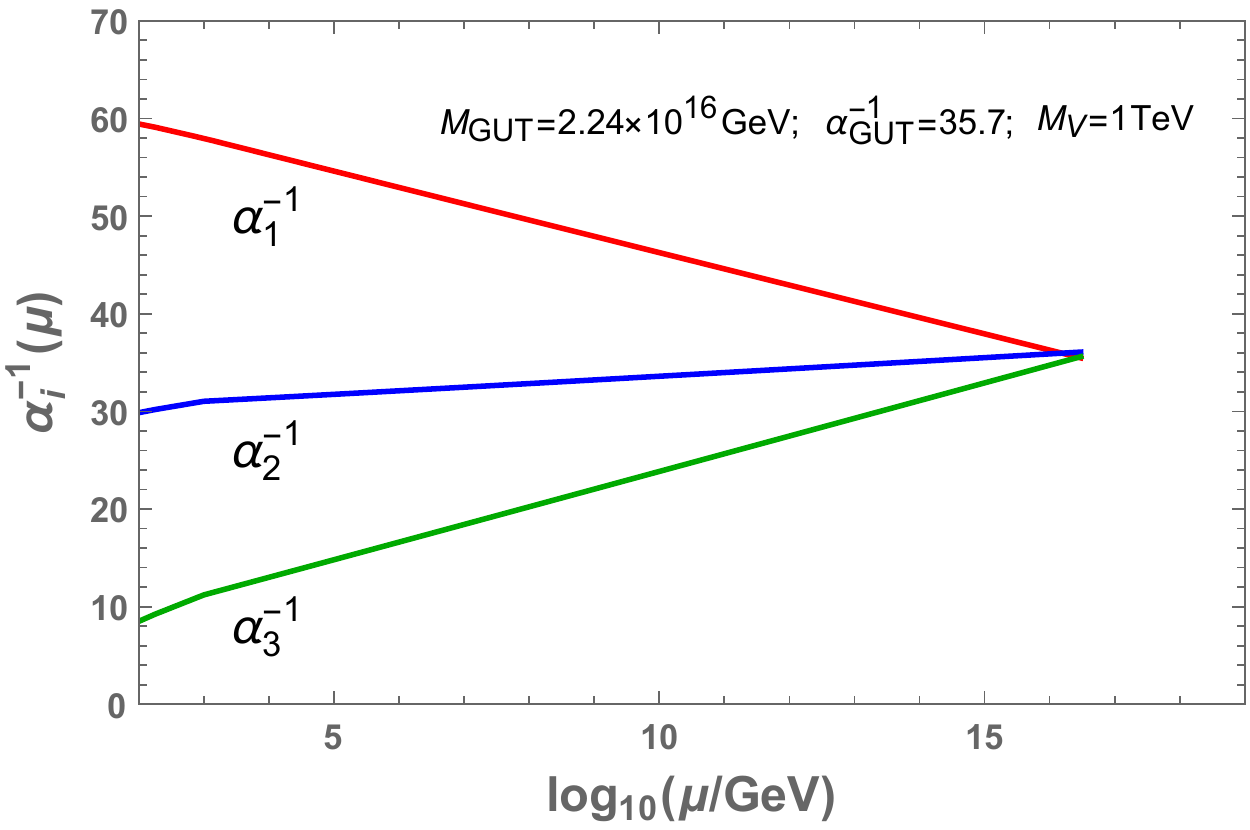}
  \caption{Two-loop gauge coupling evaluation.}
  \label{GCU}
\end{figure}

{\bf Gauge Coupling Unification~--}~
We study the gauge coupling unification by taking $M_V =1~{\rm TeV}$ and $ \mu_i =0$ in Eq.~(\ref{Yukawa}) and using
 two-loop Renormalization Group Equations (RGEs). The result is given
in Fig.~\ref{GCU}.  Defining the gauge coupling unification condition
as $\alpha^{-1}_{\rm GUT} \equiv \alpha^{-1}_{1}=( \alpha^{-1}_{2}+ \alpha^{-1}_{3})/2$,
 we obtain $\alpha^{-1}_{\rm GUT}=35.7$ and the GUT scale
$M_{\rm GUT}=2.2 \times10^{16}$~GeV. The difference between $\alpha^{-1}_{\rm GUT}$ and
$\alpha^{-1}_{2}/\alpha^{-1}_{3}$ is about $1.0\%$ or so. With the approximation formulae
in Ref.~\cite{Dutta:2016jqn}, we obtain the proton lifetime for the decay channel $p\to e^+ \pi^0$
via heavy gauge boson exchanges  around $5\times 10^{35}$ years.

{\bf Dark Energy~--}~Similar to the NMSM, we simply postulate a
cosmological constant of the observed size
\begin{equation}
  {\cal L}_\Lambda = (2.3 \times 10^{-3}~{\rm eV})^4.
\end{equation}

{\bf Neutrino Masses and Mixing and Baryon Asymmetry~--}~The neutrino masses
and mixing can be explained via seesaw mechanism~\cite{Seesaw} since the right-handed
neutrinos (and $N_x^c/N_x$) are very heavy from Eq.~(\ref{Yukawa}). Also,
 the baryon asymmetry can be explained via thermal leptogenesis~\cite{Fukugita:1986hr}. The right-handed
neutrinos are in the thermal equilibrium in the early Universe,
and the lepton asymmetry is generated from the CP violating decays of the lightest right-handed neutrino
when it is out of thermal equilibrium.
The nonperturbative sphaleron interactions violate $B+L$
but preserve $B-L$, and then the baryon asymmetry is generated
from the lepton asymmetry.

{\bf Dark Matter and Inflation~--}~To unify the dark matter particle
and inflaton, we introduce a real scalar $S$ with a $Z_2$ symmetry so that it is stable.
The potential for $S$ and $\phi$ is
\begin{eqnarray}
V&=& \lambda_{\phi} (|\phi|^2 -v^2)^2 +\frac{1}{2}m_S^2 S^2+\frac{k}{2}|\phi|^2S^2+V_I(S),~ \nonumber \\
V_I(S)&=&A~\tanh^4(S/f)~,~
\label{DMI-Potential}
\end{eqnarray}
where $m_S$ is around the electroweak scale, and $f$ is a mass parameter in the unit
of the reduced Planck scale $M_{\rm Pl}$. 
Thus, the inflaton potential is given by
$V_I(S)$, which is the $\alpha$-attractor model~\cite{Kallosh:2013yoa}.
In terms of the well-known slow-roll parameters
\begin{eqnarray}
	\epsilon=\frac{\rm{M}^2_{Pl}}{2} \left(\frac{V_I'}{V_I}\right)^2~,~
	\eta=\rm{M}^2_{Pl}\left(\frac{V''_I}{V_I}\right)~,~
	\zeta=\rm{M}^4_{Pl} \left(\frac{V'_I V'''_I}{V^2_I}\right)~,~\nonumber
\end{eqnarray}
where $X' \equiv dX/dS$, the scalar spectral index, tensor-to-scalar ratio, running of the scalar spectral index,
and power spectrum are respectively
\begin{eqnarray}
n_s&=&1-6\epsilon+2\eta~,~ r~=~16\epsilon ~,~ \nonumber \\
\alpha_s~=~\frac{\dd n_s}{\dd \ln k}&=&-24\epsilon^2+16\epsilon \eta-2\zeta~,~
P_s~=~\frac{V}{24\pi^2\epsilon}~.~
\end{eqnarray}
From the Planck, Baryon Acoustic Oscillations (BAO),
and BICEP2/Keck Array data~\cite{Ade:2015lrj,Array:2015xqh},
we have
\begin{eqnarray}
n_s=0.968\pm0.006\;,\;r= 0.028^{+0.026}_{-0.025}\;,\; \nonumber \\
\alpha_{s}=-0.003 \pm 0.007~,~P_s =2.20\times 10^{-9}~.~
\end{eqnarray}

\begin{figure}
  \centering
  \includegraphics[width=\columnwidth]{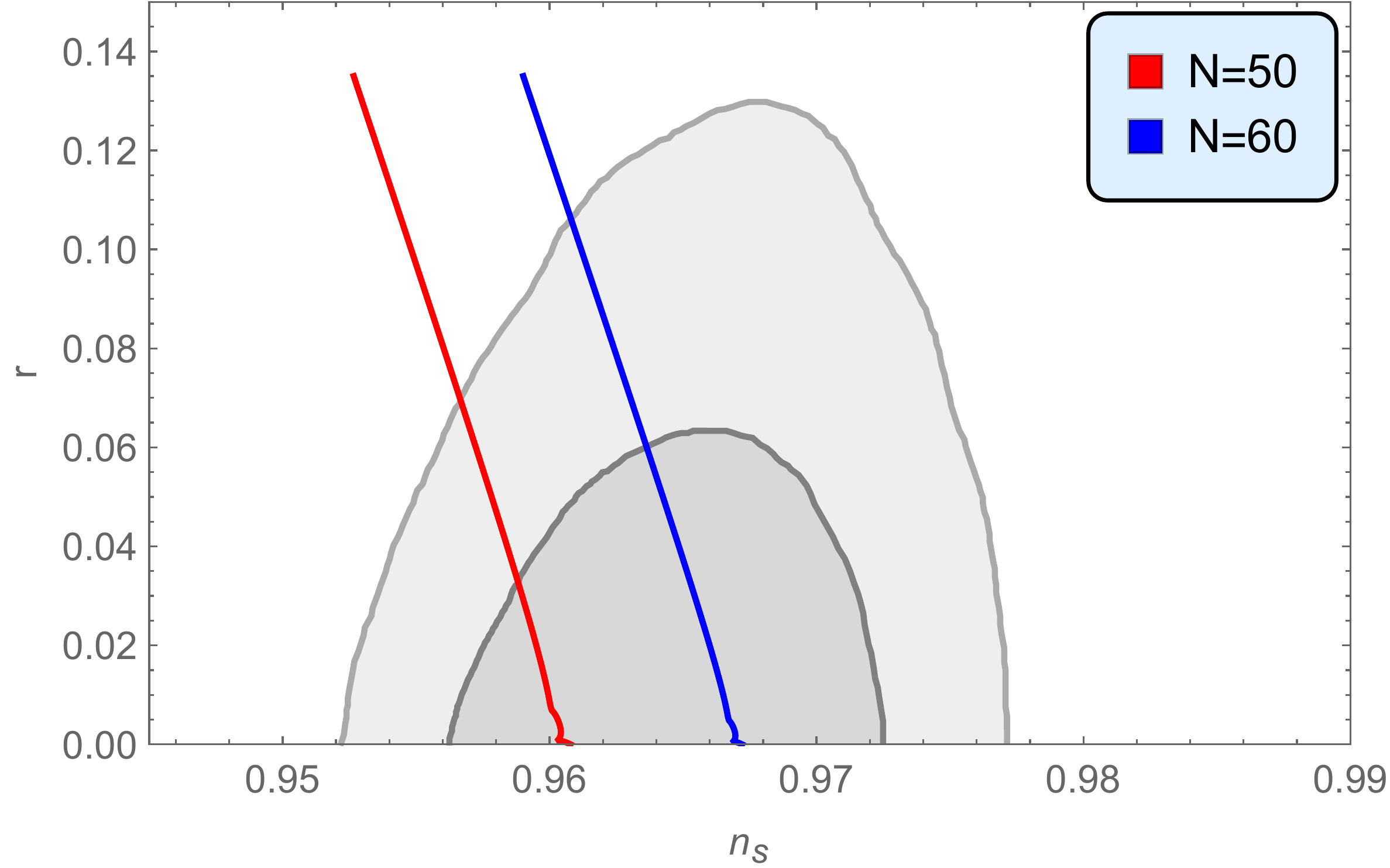}
  \caption{$n_s$ versus $r$ plots compared with Planck 2015 results~\cite{Ade:2015lrj} for TT, TE, EE $+$ lowP, at the $95\%$ CL and $68\%$ CL.}
  \label{nsr}
\end{figure}

\begin{figure}
  \centering
  \includegraphics[width=\columnwidth]{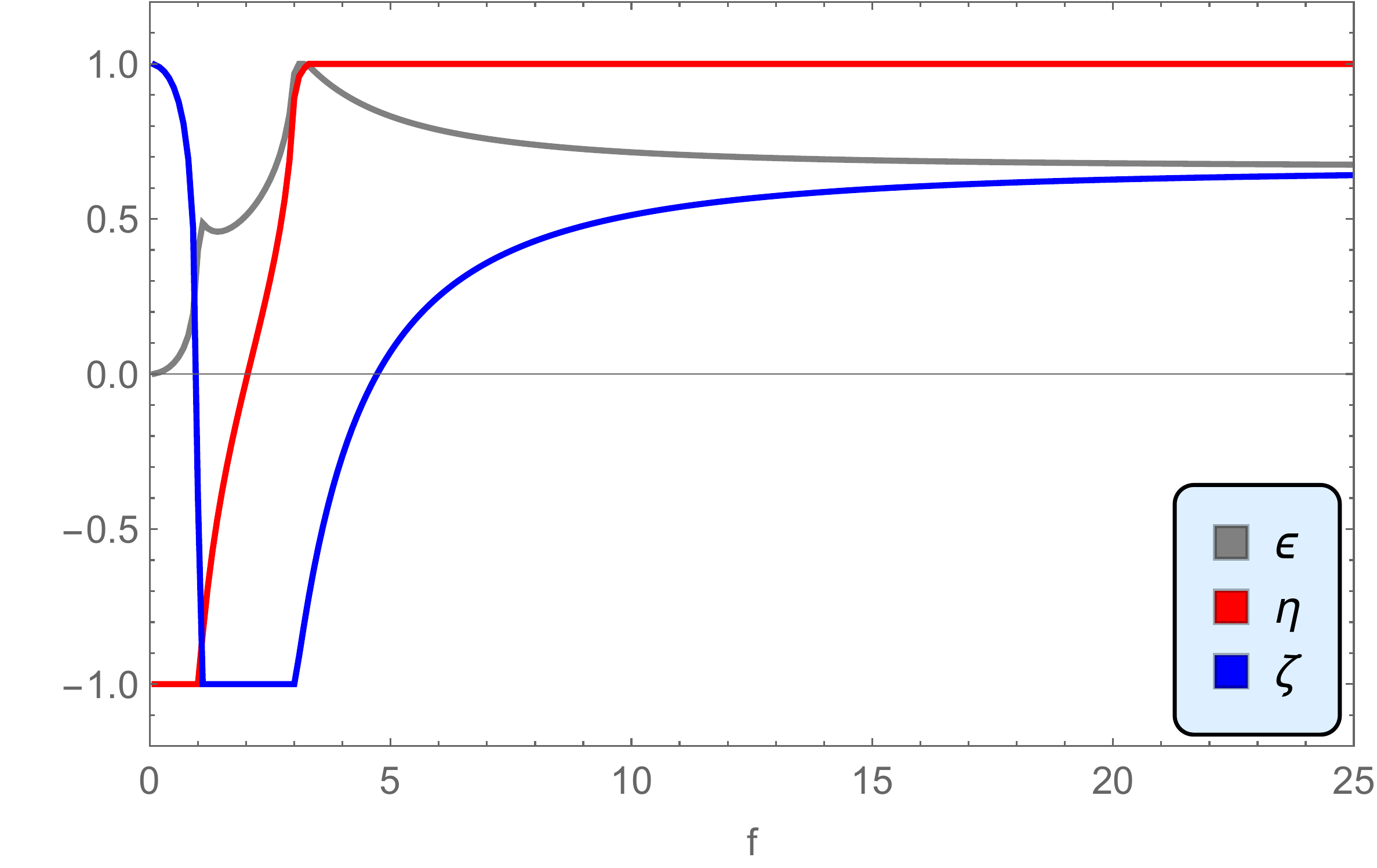}
  \caption{The slow-roll parameters $\epsilon$, $\eta$, and $\xi$ versus $f$, which is 
in the unit of $M_{\rm Pl}$. }
  \label{slowrollpara}
\end{figure}

In Fig.~\ref{nsr}, we present the numerical results for $r$ versus $n_s$, where the inner and outer
circles are $1\sigma$ and $2\sigma$ boundaries, respectively, from
the Planck 2015 results~\cite{Ade:2015lrj} for TT, TE, and EE $+$ lowP.
Therefore, our model can be highly consistent with the experimental data.
Because $f$ is the only parameter which determines the inflationary observable
$n_s$, $r$, and $\alpha_s$, we present the slow-roll parameters $\epsilon$, $\eta$, and $\xi$ versus
$f$ in Fig.~\ref{slowrollpara}. Inflation ends when any slow-roll parameter violates
the slow-roll condition.
When $f\leq 1.0~M_{\rm Pl}$ and $f\geq3.3~M_{\rm Pl}$, $\eta$ violates the slow-roll condition $|\eta|<1$.
When $1.0~M_{\rm Pl}<f\leq3.0~M_{\rm Pl}$, $\zeta$ violates the slow-roll condition $|\zeta|<1$.
And when $3.0~M_{\rm Pl}<f\leq3.2~M_{\rm Pl}$, $\epsilon$ violates the slow-roll condition $\epsilon<1$.

To have $(n_{s}, r)$  within the 1$\sigma$ and  2$\sigma$ regions of the Planck 2015 results
for TT, TE, and EE$+$ lowP in Fig.~\ref{nsr}, we obtain that
$f$ should lie in the ranges $0< f\leq13.4~M_{\rm Pl}$ for $N=60$ and $0< f\leq7.3~M_{\rm Pl}$ for $N=50$,
and in the ranges $0< f\leq 18.8~M_{\rm Pl}$ for $N=60$ and $0< f\leq 11.2~M_{\rm Pl}$ for $N=50$, respectively.
 The numerical values of $\alpha_{s}$ are always very small at the order of $10^{-4}$.
Also, the minimum of inflaton potential $V_I(S)$ is at $\phi=0$, and interestingly,
the inflaton potential will not give any mass to $S$ due to $(d^2V_I(S)/dS^2)^{1/2}|_{S=0}=0$.
Thus, after inflation, $S$ become a dark matter particle, and its Lagrangian is reduced
to that of the NMSM since $V_I(S)$ is negligible at low energy. Therefore, $S$ can be a viable dark matter candidate.
The current viable parameter space is that the dark matter mass is close to $62.5$~GeV via Higgs resonance
for small $k$ ($k \sim 0.06$ or smaller), or the dark matter mass is larger than about 450~GeV 
for relatively large $k \sim 0.2$~\cite{He:2016mls, Escudero:2016gzx}.
Let us give a benchmark point with $f=10.0~M_{\rm Pl}$. We obtain
$n_s=0.964592$, $r=0.0442495$, $\alpha_{s}=-0.0006154$ and $N=60$ for the initial value
$S_i=15.3542 ~ M_{\rm Pl}$ and final value $S_e=3.13524 ~ M_{\rm Pl}$ from the violation
of the slow-roll condition,
which fit the experimental data very well. Also, $A=2.08924\times10^{-9}~ M^4_{\rm Pl}$ can be determined
from $P_s=2.20\times10^{-9}$. We expand $V_I(S)$ at $S=0$, and get
\begin{eqnarray}
&& V_I(S)=2.0892\times 10^{-13} S^4-2.7856\times 10^{-15} \frac{S^6}{M_{\rm Pl}^2}  \nonumber \\
&& +2.5071\times10^{-17} \frac{S^8}{M_{\rm Pl}^4}
-1.8748\times 10^{-19} \frac{S^{10}}{M_{\rm Pl}^6}
+O\left(S^{12}\right).\nonumber
\end{eqnarray}
Therefore, $V_I(S)$ can indeed be neglected at low energy.

\begin{figure}
  \centering
  \includegraphics[width=\columnwidth]{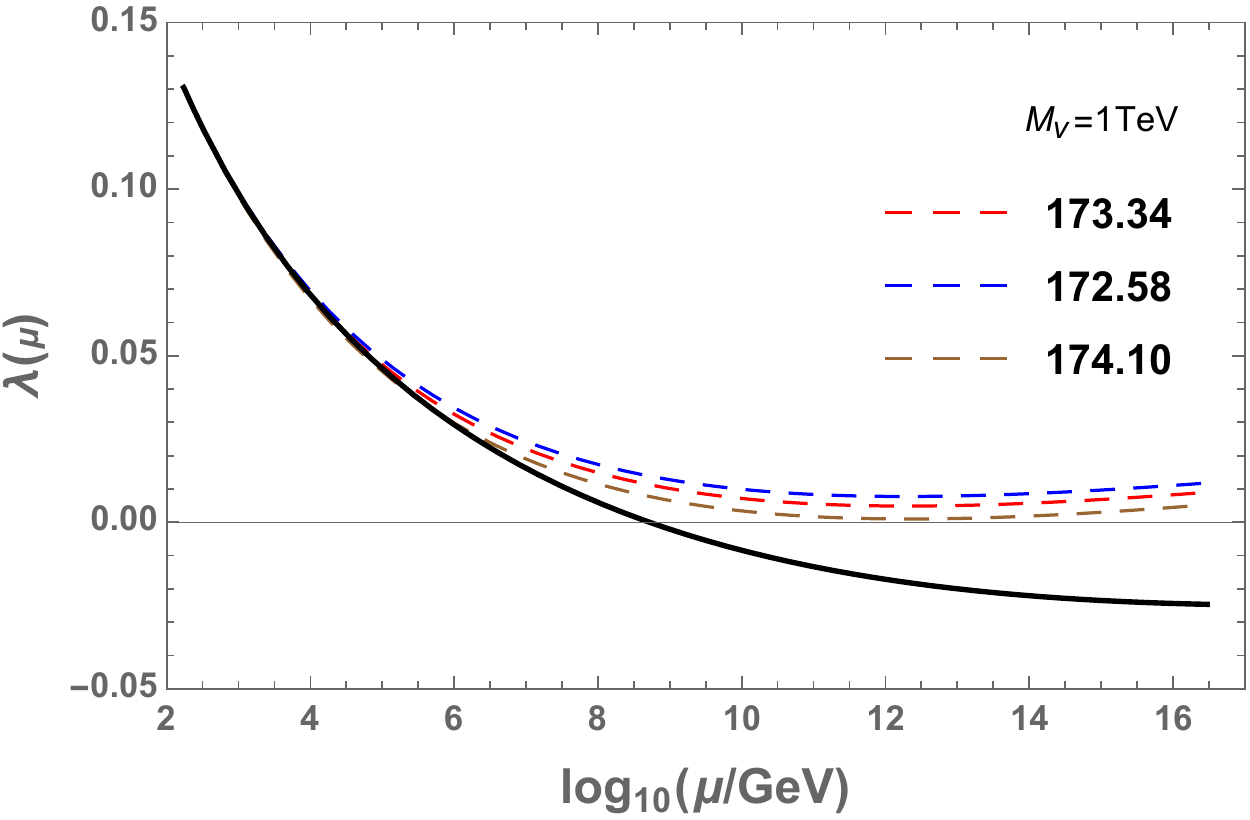}%{stability_SMNMGUT.pdf}
  \caption{The two-loop RGE evaluation for the Higgs quartic couplings.
The dashed lines stand for the Higgs quartic couplings in our model with
the  central value and one sigma deviations  of  top quark  mass.
The black solid line corresponds to the SM case with $m_t= 173.34~{\rm GeV}$.}
  \label{Stability}
\end{figure}

{\bf Stability Problem~--}~We study the two-loop RGE running of the Higgs quartic coupling.
Because it is very sensitive to the top quark mass, we consider
the central value $m_t=173.34~{\rm GeV}$ and  one $\sigma$  deviations of top quark mass ~\cite{ATLAS:2014wva}.
The numerical results are given in Fig.~\ref{Stability}.
For comparison, we also present that in the SM by taking central value of top quark mass. 
In addition, we include the dark matter contribution from the $k$ term in Eq.~(\ref{DMI-Potential})
by considering both the  small $k \sim 0.06$,
 and relatively large $k \sim 0.2$ for the viable dark matter parameter space~\cite{He:2016mls}.
We show numerically that the $k$ term can indeed be neglected.
 Similarly, the Yukawa coupling $\lambda$ in Eq.~(\ref{pgut}) between the SM Higgs 
field and GUT Higgs field can also be neglected if such coupling 
is not large, for example, 0.5, because of its short RGE running.
Therefore, to evade the stability problem, we predict the top quark mass to be smaller than
its one sigma upper bound $m_t=174.1$~GeV. The key point is that the SM gauge couplings
become stronger at high scale due to the extra vector-like particles.

{\bf Neutral Meson Mixing~--}~We would like to demonstrate that the corrections to the neutral
meson ($B^0, K^0$)  mixing in our model satisfy the strict experimental
constraints on the flavour changing neutral current processes.
Here, we assume the mass of vector-like fermions is 1 TeV.
The neutral meson mixing, like $B_{d}^{0}-\overline{B}_{d}^{0}$
and $B_{s}^{0}-\overline{B}_{s}^{0}$, is dominated by the box diagram
in which top quark and vector-like up-type quark running in the loop.
In our model, the correct values of the SM quark masses and CKM mixing
can be generated through the mixtures of vector-like quarks
with SM quarks~\cite{CGHLW-Prep}. We assume that all the elements in up-type and down-type quark mass matrices
are zero except the top quark mass.  We can use
bi-unitary transformation to diagonalize the mass matrices in up-type and down-type sectors,
and define a general $5\times  5$ non-unitary CKM matrix,
following the approach in Ref.~\cite{Branco}.
Here, we present one of realistic moduli examples of $5\times  5$ non-unitary CKM matrix ($V_{ab}$) obtained in our model 
(see Eq.~(\ref{Yukawa}))
\[\small
\left(\begin{array}{ccccc}
0.9741 & 0.2254 & 0.004109 & 1.42\cdot10^{-5} & 1.22\cdot10^{-5}\\
0.2215 & 0.995 & 0.040414 & 2.34\cdot10^{-4} & 1.87\cdot10^{-4}\\
0.008177 & 0.04004 & 1.009 & 0.01226 & 0.00981\\
2.6\cdot10^{-6} & 0.00132 & 0.380141 & 2.2\cdot10^{-6} & 1.1\cdot10^{-5}\\
0 & 0 & 0 & 0 & 0
\end{array}\right).
\]
We define $\lambda_{qq^{'}}^{a} \equiv V_{aq'}^{*}V_{aq}$ for mesons
with down quarks. The correction to the mixing $\left(M_{12}\right)_{qq^{'}}$ which is the 12 element of $2\times  2$ mass matrix in the neutral meson oscillation system is
\begin{eqnarray*}
\frac{\left(M_{12}\right)_{qq^{'}}}{\left(M_{12}^{SM}\right)_{qq^{'}}}  &\sim& 1+\left(\frac{\lambda_{qq^{'}}^{U}}{\lambda_{qq^{'}}^{t}}\right)^{2}\frac{S\left(x_{U}\right)}{S\left(x_{t}\right)}+2\frac{\lambda_{qq^{'}}^{U}}{\lambda_{qq^{'}}^{t}}\frac{S\left(x_{U},\:x_{t}\right)}{S\left(x_{t}\right)},
\end{eqnarray*}
where $qq^{'}$ stands for quarks participating in the box diagram leading to the neutral meson mixing \cite{Branco}. $S\left(x_{t}\right)$ and $S\left(x_{U},\:x_{t}\right)$ are the
IL functions defined as in Ref.~\cite{Silva}, $x_{t}=\left(m_{t}/M_{W}\right)^{2}$
and $x_{U}=\left(m_{U}/M_{W}\right)^{2}$. Following their convention,
the corrections with respect to the SM predictions are defined as
$
\Delta\left(P_{0}\right)\equiv \left|\left({M_{12}}/{M_{12}^{SM}}\right)_{P_{0}}\right|-1~,
$
where $P_{0}$ could be $K^{0}$, $B_{d}^{0}$ and $B_{s}^{0}$. Using
the $5\times5$ CKM matrix shown above, we can obtain the corrections
with respect to the SM predictions in our model
$\left(\Delta\left(K^{0}\right),~ \Delta\left(B_{d}^{0}\right),~
\Delta\left(B_{s}^{0}\right)\right) = \left(
5.5\cdot10^{-5},~
6.3\cdot10^{-4},~
0.066\right)$.
Thus, the corrections to $\Delta M_{K^{0}}$ and $\Delta M_{B_{d}^{0}}$
are very small compared with the SM predictions. However, the correction
to $\Delta M_{B_{s}^{0}}$ 
in our model is 6.6\% which cannot be neglected.

{\bf  Comments on Reheating~--}~The challenge question for our inflaton and dark matter
unification scenario is reheating.
There are two kinds of solutions: (i) The inflaton
decay only occurs during the initial stage of field oscillations
after inflation, and then is kinematically forbidden at late time~\cite{Bastero-Gil:2015lga}.
In this approach, we need to introduce two SM singlet fermions, and then it is not minimal;
(ii) The $Z_2$ symmetry is broken at high scale at a meta-stable vacuum and thus
 inflaton can decay for reheating. After the meta-stable vacuum decays into the real vacuum,
the $Z_2$ symmetry is restored, and then inflaton can be a dark matter candidate.
Because the first solution has already been studied previously~\cite{Bastero-Gil:2015lga}, 
we will not repeat it here.
Thus, we shall briefly explain the idea for the second solution~\cite{CGHLW-Prep}.
In this solution, we consider the following inflaton potential $V_I(S)$
\begin{eqnarray}
V_I = \left\{\begin{array}{lll}
A~\tanh^4(S/f)~& \hbox{for} & |S| > S_b~{\rm and }~ |S| < S_a \\
\Lambda_S + \frac{\lambda_S}{2} (S^2-v_S^2)^2 ~& \hbox{for} &  S_a < |S| < S_b
\end{array}\right. , \nonumber
\end{eqnarray}
where $0< S_a < v_S < S_b < S_e$. To have the continuous inflaton potential, we require 
\begin{eqnarray}
A~\tanh^4(S/f) = \Lambda_S + \frac{\lambda_S}{2} (S^2-v_S^2)^2~,~
\end{eqnarray}
at $|S|= S_a$ and $|S| = S_b$.
Thus, $\langle S \rangle =v_S$ is a meta-stable vacuum, and the $Z_2$ symmetry is broken at this vacuum.
With $\lambda_S > k$, we have $m_S > 2 m_\phi$ at the meta-stable vacuum, and
 $S$ can decay into two Higgs particles. Thus, we can indeed realize the reheating. 
Moreover, we choose the proper parameters $\Lambda_S$, $\lambda_S$, and $v_S$
so that the meta-stable vacuum can decay into the real vacuum with $\langle S \rangle=0$ just after
reheating. Thus, the $Z_2$ symmetry will be restored, and $S$ can be a dark matter
candidate as well.
The detailed study will be given elsewhere~\cite{CGHLW-Prep}.

{\bf  Discussions and Conclusion~--}~We have proposed the non-supersymmetric minimal
GUT with flipped $SU(5)\times U(1)_X$ gauge symmetry
and one-pair of vector-like particles, which can incorporate all
the convincing new physics beyond the SM
based on the principle of the minimal particle content.
The gauge coupling unification can be realized,
proton lifetime is about $5\times 10^{35}$ years,
and the doublet-triplet splitting problem at tree level as well as stability problem
can be solved. The possible signals from neutral meson mixing have been studied as well.
 Remarkably, we proposed a brand new scenario for the
unification of inflaton and dark matter particle,
which can be applied to the generic scalar dark matter models.

{\bf Acknowledgments~--}~This research was supported in part
by the Projects 11605049, 11475238 and 11647601
supported by National  Science Foundation of China,
and by Key Research Program of Frontier Sciences, CAS. The
work of IG is supported in part by Bartol Research Institute.

%%%%%%%%%%%%%%%%%%%%%%%%%%%%%%%%%%%%%%%%%%%%%%%%%%%%%%%%%%%%%%%%%%%%%%%%%%%%

\end{document}